# Electron Impact Ionization Close to the Threshold: Classical Calculations

**F Sattin[1#] and K Katsonis[2&]**

[1] Consorzio RFX-Associazione Euratom-Enea sulla Fusione, Corso Stati Uniti 4, 35127 Padova, Italy

[2] Laboratoire de Physique des Gaz et des Plasmas, Universitè Paris XI, 91405 Orsay, France

**Abstract**

In this paper we present Classical Trajectory Monte Carlo (CTMC) calculations for single and multiple electron ionization of Argon atoms and ions in the threshold region. We are able to recover the Wannier exponents $\alpha$ for the power-law behavior of the cross section $\sigma$ versus excess energy: the exact value of the exponent as well as the existence of its saturation for multiple ionization appear to be related to how the total binding energy is shared between target electrons.

[#] sattin@igi.pd.cnr.it
[&] konstantinos.katsonis@lpgp.u-psud.fr

A detailed modelization of atomic processes involving rare gas atoms is needed for characterization of low-temperature plasmas such as those encountered, e.g., in the scrape-off layer of magnetic fusion devices and in plasma thrusters for astrophysical applications. Depending on the plasma conditions and on the accuracy sought, even a zero-dimension modeling of such plasmas necessitates the formulation of a full collisional-radiative model or one of its simplifications (coronal-radiative or even a coronal model). In so doing, among the numerous needed data, electron collision ionization rate coefficients are of paramount importance whenever we are interested to mostly ionizing plasmas (Drawin and Katsonis 1981, Janev and Katsonis 1987).

This paper presents a theoretical study of the single and multiple electron ionization of Argon in several ionization stages at impact energy close to the threshold. It is intended as a part of a broader study aimed to produce ionization cross sections over a larger energy range, suitable for the computation of rate coefficients (Katsonis *et al.* 2002a, 2003); however we think that this investigation deserves an interest *per se*. In fact, the threshold breakup of complex systems into several charged components has been a subject of interest at least since the paper by Wannier introducing its famous threshold law $\sigma \propto \Delta E^{\alpha}$ (Wannier 1953) based on analytical considerations on the three-body problem; work on two-body threshold breakup goes back till Wigner (1948). Many theoretical and experimental studies have appeared since then, trying to validate or improve that law. Basically, it has been confirmed, but research still focuses on the precise evaluation of exponent $\alpha$. An updated bibliography can be found in Gstir *et al.* (2002).

In this work we shall compute ionization cross sections by direct numerical integration of the equations of motion for a few-body atomic system consisting of one projectile electron, *n* target electrons initially bound to one nucleus, and the remaining electronic cloud after the removal of the *n* electrons together with the atomic nucleus. This procedure constitutes the basis of the well-known CTMC method, widely applied to the study of collisions between heavy particles (Abrines and Percival 1966a,b). The CTMC method has been applied also to collisions involving light projectiles, although with some concern dictated by the neglecting of quantum mechanical features, thought to be important in this case (Schultz *et al*. 1992). However, Wannier studies (see also Rost 1994) suggest that the main features of the ionization process, even near the threshold energy, are still retained by limiting to the classical picture in accordance with the extension of the CTMC application to low energies as discussed in Katsonis and Varvoglis (1995).

The analytical approach, adopted by Wannier and followers, allows to have a physical insight of the dynamics of the process, but it does involve several approximations, so as to make one wonder how much accurate is the quantitative estimate of $\alpha$. The present numerical approach is elucidating this point, since the equations solved here are exact. Further to preliminary investigations corroborating the Wannier $\alpha$ value for simple ionization (Katsonis *et al.* 2002b), we address here the question of the value of $\alpha$ for multiple ionization, which is still a matter of discussion (see Gstir *et al.* 2002).

The precise estimate of threshold laws is more than of academic interest: quite recently, Pattard and Rost (1999) suggested that most ionization cross sections can be interpolated with fair accuracy by a simple analytical parameterization, whose form is simply taken as the product of a low-energy part (given by the Wannier law) and a high-energy one, given by a modified Born law where the logarithm is discarded (thus, $\sigma \propto 1/E$). It is outside the scope of the present paper to explain how the logarithmic term must be recovered for high energies, hence we refer the reader to Katsonis *et al.* (2002a). One can adapt the scaling function to any particular process by simply assigning the value of a few parameters: in the form given by Pattard and Rost these parameters are the value of the maximum of the cross section and the energy at which this maximum occurs. Reducing the computation of the cross section over the whole energy range to the estimating only of these parameters would be, of course, a major improvement in the cases where a large amount of these data is needed. Most recently, a numerical two parameter formula based on the Wannier threshold and in accordance with the Born approximation has been devised (Katsonis *et al.* 2003).

It is well known that the CTMC method does not handle easily multi-electron systems, since electron-electron interactions cannot be treated satisfactorily within the classical framework. Nevertheless, the independent electron approximation is usually done, in which the target electrons interact with the nucleus and the projectile, but not between them. Here we are interested to energies near the threshold, where we will resort to this approximation. For Argon, as well as for other heavy rare gas elements, this appears to be not too crude an approximation (Katsonis *et al.* 2002a).

The bound electron position and velocity are initialized according to the microcanonical distribution. The number of target electrons varies with the number of ionizations; thus we have initialized one electron when we wished to study single ionization, two for double ionization, etc.... The results must then be multiplied by the number of active electrons to obtain absolute cross sections. The mass of Argon nucleus is so large in comparison with those of the light particles that it stays practically motionless during the scattering. Its equations of motions are therefore not solved. In the CTMC method the probability for the occurrence of a process is identified with the fraction of useful trajectories. This fraction is a decreasing function of the number of simultaneous ionizations, of the nuclear charge and of the projectile velocity; therefore, the least probable processes need a huge number of runs. In this work we do not carry on an investigation of all ionization processes, and present only results for single and double ionization. An extended set of investigations was carried on at higher velocities, including triple ionization and charge states up to Neon-like $Ar^{8+}$. Its results will be presented elsewhere (Katsonis *et al.* 2003).

An interesting point is raised in connection with the choice of the binding energy of the electrons: let $V_n$ be the *total* energy needed for ionizing *n* electrons. Two choices are possible: the former is to give the first electron a binding energy $V_1$, the second $V_2 - V_1$ and so on. This means that one has both tightly and weakly bound electrons. The second choice treats instead all electrons on equal footing and assigns a binding energy $V_n/n$ to each of them. It has been pointed out previously (Katsonis *et al.* 2002a) that whenever the valence electrons are part of the same configuration the latter option is prevailing, being closer to some theoretical models of ionization based on statistical distribution of projectile energy. Both options have been investigated here. Ionization potentials were taken from (Moore 1949); they can also be found in table 2 of Gstir *et al.* (2002).

Threshold laws are, by definition, valid only over small energy ranges. This is not a problem for analytical calculations, where the limit to zero can always be taken, but raises a serious question for numerical or experimental work. The papers of Rost (1994), Koslowski *et al.* (1987), and Gstir *et al.* (2002) have shown that a safe choice for the width of the threshold region is $1 < E/V_n < 1.1\sim1.2$.

There is presently little controversy about threshold exponents for the single ionization process: Wannier theory predicts $\alpha = 1.127$ for neutral atoms, slowly decreasing towards 1 for highly charged ions. Recent experiments (Gstir *et al.* 2002) tend to give a slightly larger value, closer to 1.2. In figure 1 we plot our results for single ionization of $Ar^0$ and $Ar^{5+}$. Both curves can be nicely fitted with a power law: the exponents are reproduced within the plots and well agree with Wannier exponents (the Wannier exponent for single ionization from $Ar^{5+}$ is $\alpha \sim 1.017$). The error bars on these exponents as given by the fitting routine are rather small, but there is a larger uncertainty due to the precise definition of the threshold region (see the discussion in the previous paragraph). It is interesting to notice about this, that $Ar^{5+}$ seems to have a larger threshold region than $Ar^0$: if we should include higher energy points in the $Ar^0$ fit, the computed exponent would decrease noticeably (around 1.11 by increasing the range up to $E/V = 1.2$). On the contrary, $\alpha$ does not vary appreciably if we skip the last point in the fitting of $Ar^{5+}$.

Double ionization is a more debated argument. Classical phase space arguments predict that $\sigma \propto (\Delta E)^n$, with $n$ ($> 1$) number of ejected electrons: see the half-page paper (Wannier 1955), the statistical approach of Russek (1963), or Koslowski *et al.* (1987). Experiments suggest that this law is valid for double ionization, but is increasingly less satisfied with the number of ionizations, the correct exponent being lesser than $n$ (Koslowski *et al.* 1987, Gstir *et al.* 2002). Actually, there appears to be a saturation for $n > 3\sim4$. A value $\alpha < n$ is predicted by (Lebius *et al.* 1989, 1991). On the other hand, values even larger than $n$ have been predicted by some other models (Klar and Schlecht 1976, Grujic 1983).

In figure 2 we show our results for double ionization of $Ar^0$, together with the power-law fitting. It is remarkable how the two different choices for the initial electron distribution affect the results. When the electrons are treated all as equal, the $\alpha = n$ law is recovered rather nicely. On the other hand, when they are assigned different ionization potentials, the single-ionization law is found with a good approximation. This result admits a straightforward interpretation, provided the two electrons have fairly different ionization potentials: the double ionization process is divided into two successive independent single collisions. In the former, the projectile knocks off the least bound electron. This happens with a high probability, since ionization probability increases with velocity and the projectile has not yet lost its energy. The actual frequency is therefore regulated by the least probable process - the inner shell ionization -, which follows the single-ionization power law. Indeed, we suggest that this mechanism is perhaps responsible for the saturation of $\alpha$ at triple or higher ionization, albeit this must not be considered a rigorous demonstration, rather a working hypothesis to be possibly confirmed by further more refined analysis.

Our CTMC calculations confirm once more what has been stated in the past, namely that the threshold electron impact process can be adequately described within a purely classical framework. As for multiple ionization, it is shown that they can support the prediction by the simplest phase space arguments, namely $\alpha \sim n$. This will be useful for the prosecution of this work, i.e. the fitting of numerical cross sections with the analytical function by Pattard and Rost (1999).

**Figures**

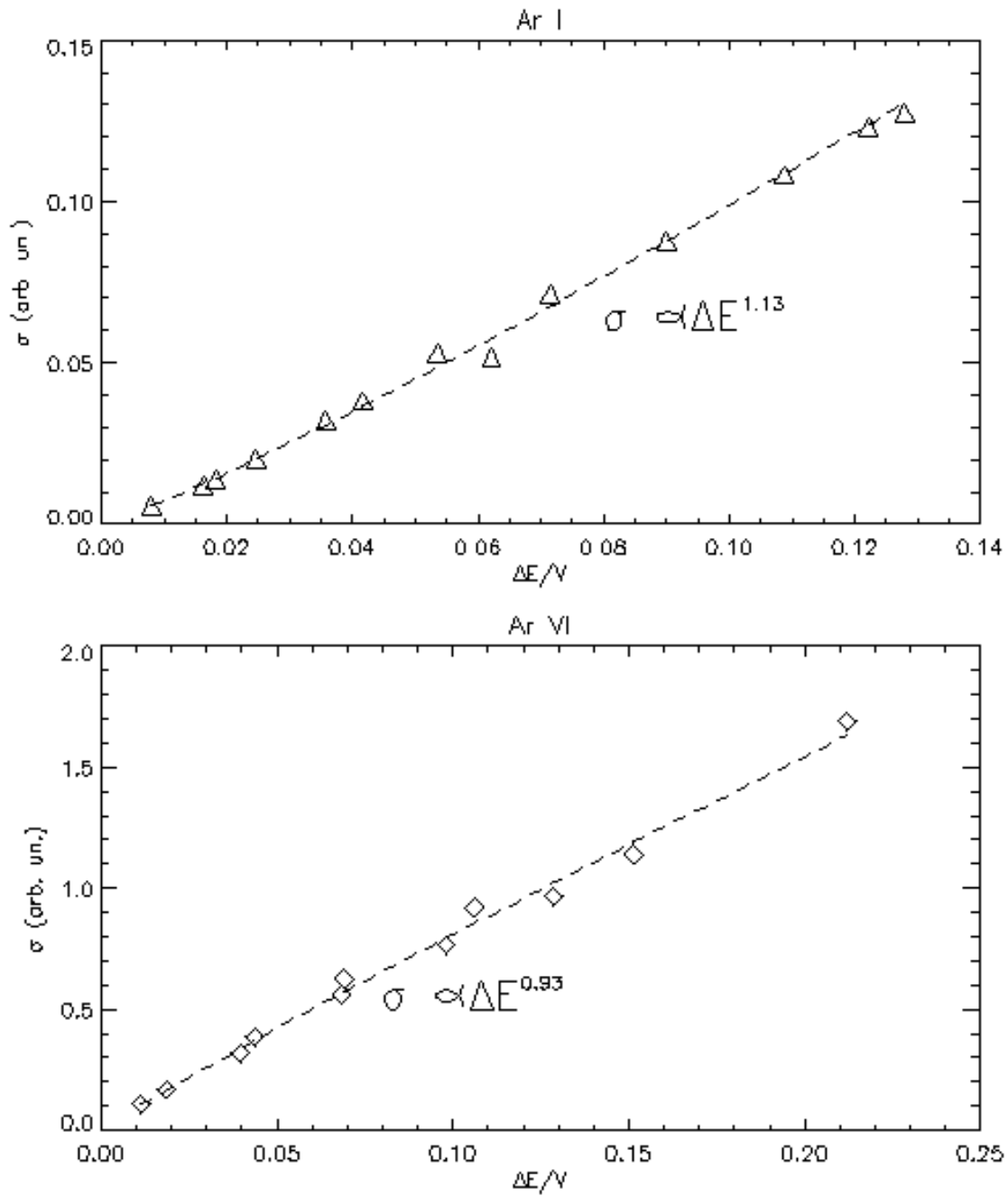

Fig. 1. Single ionization of Argon (in arbitrary units) versus normalized excess energy. Top, ionization of Ar I; bottom, ionization of Ar VI. Fitting power-law curves are also shown.

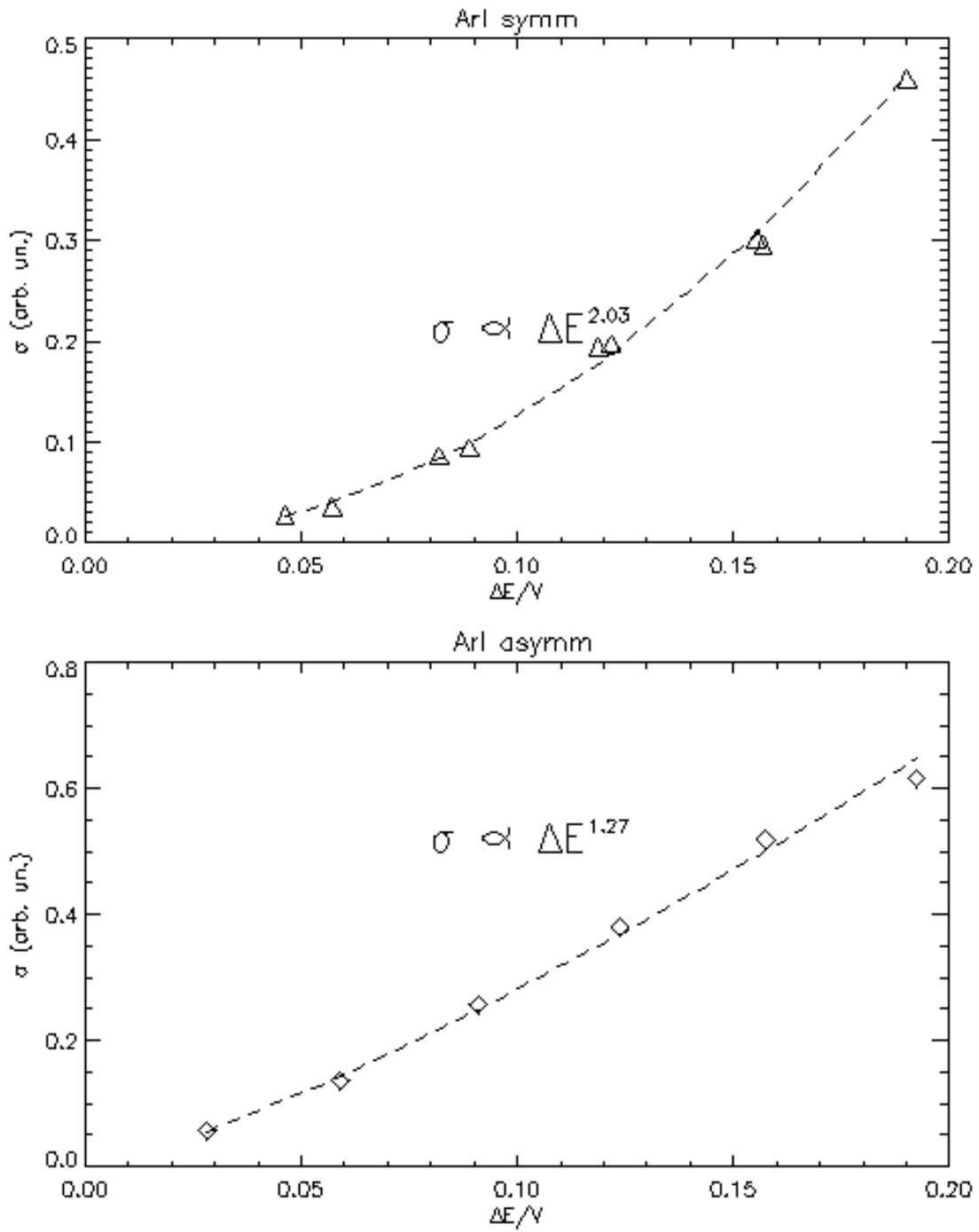

Fig. 2. Double ionization of neutral Argon (in arbitrary units) versus normalized excess energy. Top, both electrons have the same energy, equal to half the total ionization potential *V*. Bottom, the two target electrons are given different ionization potentials. Fitting power-law curves are also shown.